# Tunable hyperbolic Landau-level polaritons in charge-neutral graphene nanoribbon metasurfaces

*Kateryna Domina*[†§], *Tetiana Slipchenko*[¤], *D.-H.-Minh Nguyen*[†#], *Alexey B. Kuzmenko*[‡], *Luis Martin-Moreno*[∥⊥], *Dario Bercioux*[†¶], *Alexey Y. Nikitin*[†¶*]

[†] Donostia International Physics Center, 20018 Donostia-San Sebastian, Spain

[§] Universidad del Pais Vasco/Euskal Herriko Unibertsitatea, 48940 Leioa, Spain

[¤] Instituto de Ciencia de Materiales de Madrid (ICMM), CSIC, 28049 Madrid, Spain

[#] Advanced Polymers and Materials: Physics, Chemistry and Technology, Chemistry Faculty (UPV/EHU), 20018 San Sebastian, Spain

[‡] Department of Quantum Matter Physics, University of Geneva, 1211 Geneva, Switzerland

[∥] Instituto de Nanociencia y Materiales de Aragon (INMA), CSIC-Universidad de Zaragoza, 50009 Zaragoza, Spain

[⊥] Departamento de Fisica de la Materia Condensada, Universidad de Zaragoza, 50009 Zaragoza, Spain

[¶] IKERBASQUE, Basque Foundation for Science, 48009 Bilbao, Spain

ABSTRACT. Magnetized charge-neutral graphene supports collective hybrid electronic excitations - polaritons - which have a quantum origin. In contrast to polaritons in doped graphene, which arise from intraband electronic transitions, those in charge-neutral graphene originate from interband transitions between Landau levels, enabled by the applied magnetic field. Control of such quantum polaritons and shaping their wavefronts remains totally unexplored. Here, we design an artificial two-dimensional quantum material formed by charge-neutral graphene nanoribbons exposed to an external magnetic field. In such a metasurface, quantum polaritons acquire a hyperbolic dispersion. We find that the topology of the isofrequency curves of quantum hyperbolic magnetoexciton polaritons excited in this quantum material can change, so that the shape of the isofrequency curves transforms from a closed to an open one by tuning the external magnetic field strength. At the topological transition, we observe canalization phenomena, consisting of the propagation of all the polaritonic plane waves in the continuum along the same direction when excited by a point source. From a general perspective, our fundamental findings introduce a novel type of actively-tunable quantum polaritons with hyperbolic dispersion and can be further generalized to other types of quantum materials and polaritons in them. In practice, quantum hyperbolic polaritons can be used for applications related to quantum sensing and computing.

## INTRODUCTION

In pristine graphene, close to the charge neutrality point, low-energy carriers exhibit a linear energy dispersion and a vanishing Fermi surface. However, once graphene is doped, the Fermi level shifts away from the Dirac point, the Fermi surface emerges at the two valleys, enabling electronic intraband transitions that promote graphene plasmon polaritons - hybrid coupled excitations of photons and oscillations of Dirac quasiparticles in graphene. High levels of graphene doping (up to $E_f \sim 1$ eV) can be achieved by electrostatic gating,[1,2] chemical modification,[3,4] or optical pumping,[5] which allows tuning of graphene plasmon polaritons in the terahertz to mid-infrared frequency range.[1] Additionally, in the presence of an external static magnetic field, free carriers in doped graphene experience cyclotron motion and occupy discrete Landau levels, which modify graphene plasmon polaritons dispersion and enable magneto graphene plasmon polaritons, [6–9] which allows additional chiral control of light-matter interaction[10] and quantum light generation.[11] However, despite the possibility of magnetic tuning, the electron scattering rate in the presence of a magnetic field increases,[12] which limits the magneto graphene plasmon polaritons quality factor and thus the performance of potential optoelectronic devices based on magnetized graphene.

Magnetized charge-neutral graphene (CNG) could be an alternative base for superior graphene polaritons performance due to a lower electron scattering rate. Whilst the giant intrinsic photoresponse of CNG,[13] and strong magneto-optical response of magnetized CNG[14] have been demonstrated, studies on polaritons in CNG remain rather limited. Even for small values of magnetic fields, electronic states in CNG quantize into discrete Landau levels. In this scenario, electrons in magnetized CNG experience high-energy interband (inter-Landau-level) transitions that facilitate interband quantum magnetoexciton polaritons,[15] which only recently were revealed experimentally at mid-infrared frequencies.[16,17]

Until now, quantum hyperbolic polaritons have not been reported. These are quantum polaritons with hyperbolic-shaped isofrequency curves (IFCs) - cross-sections of the dispersion in momentum space at a fixed frequency, representing all allowed in-plane wavevectors for a given mode. They stem from discrete interband transitions between Landau levels at charge neutrality. In this work, we fill this gap by introducing quantum hyperbolic magnetoexciton polaritons (QHMEPs). We show that the latter can exist in a metasurface that consists of nanoribbons of CNG placed in an external static magnetic field. An attractive feature of these QHMEPs is their magnetic-field tunability, which enables control over the shape of their IFCs. In particular, one can realize a topological transition of IFCs from close (elliptic-like) to open (hyperbolic-like) shapes. This tunability enables access to the so-called polariton canalization regime, in which the polariton's energy flows along a single direction (in practice, forming a narrow channel).

RESULTS AND DISCUSSION

To study QHMEPs, we designed a metasurface that consists of periodically placed CNG nanoribbons of width $w = 120$ nm (this value selected for illustrative purposes) aligned with the $x$-axis, as shown schematically in **Figure 1a**. For simplicity, we consider a free-standing metasurface (surrounded by air). In the absence of an external magnetic field, only modes arising from quantum confinement and edge-state structures may exist in CNG nanoribbons (see Supplementary Note S1). However, the low carrier density significantly limits the formation of polaritons. When the metasurface is subjected to a perpendicular static magnetic field $\boldsymbol{B}$, the interband transitions between quantized Landau levels in the nanoribbons are enabled, facilitating the formation of QHMEPs. The magnetic field strength can be tuned to manipulate the dispersion and topology of QHMEPs. To visualize the polaritonic response of the metasurface, we consider a vertically oriented electric dipole source $\boldsymbol{d}$, placed $50$ nm above the structure and numerically model the resulting excited out-of-plane electric field component ($E_z$).

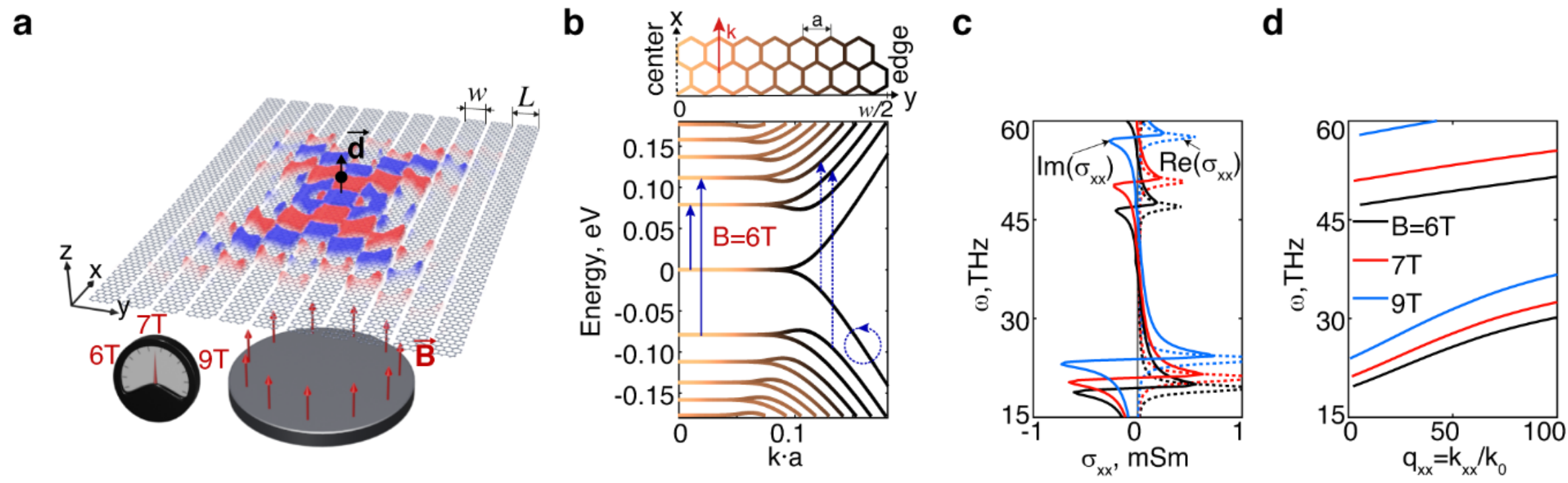

**Figure 1**. Conductivity of the magnetized CNG nanoribbons forming the metasurface. **(a)** Schematics of the studied quantum hyperbolic metasurfaces in an external static magnetic field $\boldsymbol{B}$. QHMEPs are excited by a vertically-oriented point dipole with its dipolar moment $\boldsymbol{d}$, placed above the metasurface that consists of nanoribbons of width $w$ placed periodically with period $L$. **(b)** Electronic dispersion of a 120-nm graphene armchair nanoribbon subjected to a perpendicular magnetic field. The blue arrows denote the optical transitions, $k$ is the electron momentum along the nanoribbon, $a$ is a graphene lattice constant. **(c)** Real $\mathrm{Re}(\sigma_{xx})$, and imaginary $\mathrm{Im}(\sigma_{xx})$, parts of the optical conductivity of a single magnetized CNG nanoribbon (along the nanoribbon), dashed and solid lines, respectively, as a function of frequency $\omega$, for different values of the applied magnetic field. **(d)** Estimation of the dispersion relations for polaritons in magnetized CNG nanoribbons, calculated using a simplified relation for an infinite isotropic two-dimensional (2D) conducting sheet with the isotropic conductivity $\sigma$, set to $\sigma = \sigma_{xx}$.

The polaritonic response of the metasurface depends on the optical properties of electrons residing in each graphene nanoribbon. In our work, we consider graphene nanoribbons with armchair edge structure, as zigzag edge quenches the plasmons and thus these nanoribbons have a higher polaritonic decay rate (see Supplementary Note S3).[18] Without an external field, the electronic spectrum of a graphene nanoribbon is quantized into discrete bands due to the confinement effect and disperses along the $x$-direction. The application of a perpendicular magnetic field tends to force electrons at the centre of the nanoribbon into cyclotron orbits and those at the edges into skipping orbits.[19] As shown in **Figure 1b,** this leads to the band flattening near $k = 0$, forming the Landau levels. In the central area of the nanoribbon, each Landau level is twofold degenerate because the K and K$'$ valleys coincide and are effectively decoupled. When approaching the ribbon boundaries, however, this valley degeneracy is lifted: the armchair boundary conditions enforce intervalley mixing and impose an asymmetric effective confinement potential on the electrons, causing the Landau levels to split into two branches.[20] The corresponding electronic states localize at the nanoribbon centre (depicted by yellow colour) and form chiral modes at the nanoribbon edges (depicted by black colour). Within the linear response theory, the optical transitions vary across each nanoribbon: while the Landau levels only permit interband transitions,[21] the chiral edge modes allow both interband and intraband transitions,[6,22,23] as indicated by the blue solid and dashed arrows, respectively (also, see Supplementary Note S1). Describing the electrons in an armchair nanoribbon by a tight-binding Hamiltonian with only nearest-neighbor hoppings and using the Kubo-Greenwood formula (see Methods), we obtain the anisotropic optical 2D conductivity $\hat{\boldsymbol{\sigma}}$ - $2 \times 2$ tensor, which characterizes each nanoribbon. Its component along the nanoribbon, $\sigma_{xx}$, is shown in **Figure 1c** as a function of frequency (also, see Supplementary Note S2 for other components of the conductivity tensor $\hat{\boldsymbol{\sigma}}$). The real $\mathrm{Re}(\sigma_{xx})$, and imaginary $\mathrm{Im}(\sigma_{xx})$, parts of the conductivity are shown with dashed and solid lines, respectively. Each interband transition between the Landau levels manifests at specific frequencies, at which we see pronounced peaks in $\mathrm{Re}(\sigma_{xx})$, while, $\mathrm{Im}(\sigma_{xx})$, changes its sign (according to Kramers–Kronig relation). The existence of polaritons in graphene requires the imaginary part of the conductivity to be positive, $\mathrm{Im}(\sigma_{xx}) > 0$, thus each transition forms a narrow THz frequency band where polaritons are supported. Notably, the spectral positions and amplitudes of the peaks in the conductivity spectra are tunable via the applied magnetic field (**Figure 1c**). Next, to estimate the dispersion of QHMEPs that could be supported by our metasurface, we use a simplified relation for infinite isotropic 2D conducting sheet $q = \sqrt{1 - 1/\alpha^2}$, where $q = k/k_0$ is a refractive index of the polaritonic mode with $k_0 = \omega/c$ being the free-space wavevector, and $\alpha = (2\pi/c)\sigma$ - a normalized conductivity.[24] We substitute the calculated conductivity of a single CNG nanoribbon along the ribbon ($x$-direction), $\sigma_{xx}$, into this expression. The resulting dispersions, shown in **Figure 1d**, exhibit large refractive index values $q$, corresponding to a strong confinement of the modes. However, this simplified estimation does not take into account the effect of the metasurface, i.e., the electromagnetic coupling between nanoribbons via supported polaritonic modes. The coupling between the nanoribbons is known to reshape the dispersion of polaritons strongly. In particular, it enables access to regimes not accessible in continuous graphene, such as hyperbolic dispersion and canalization,[25,26] which have not been reported for quantum polaritons yet. The conductivity of the nanoribbons calculated within the used method (see Methods) strongly depends on their width (see Supplementary Note S3). For narrower ribbons, quantum effects become more pronounced and play a significant role in shaping the properties of the supported polaritons, particularly influencing the hyperbolic dispersion and propagation losses of the polaritonic mode (see Supplementary Note S6).

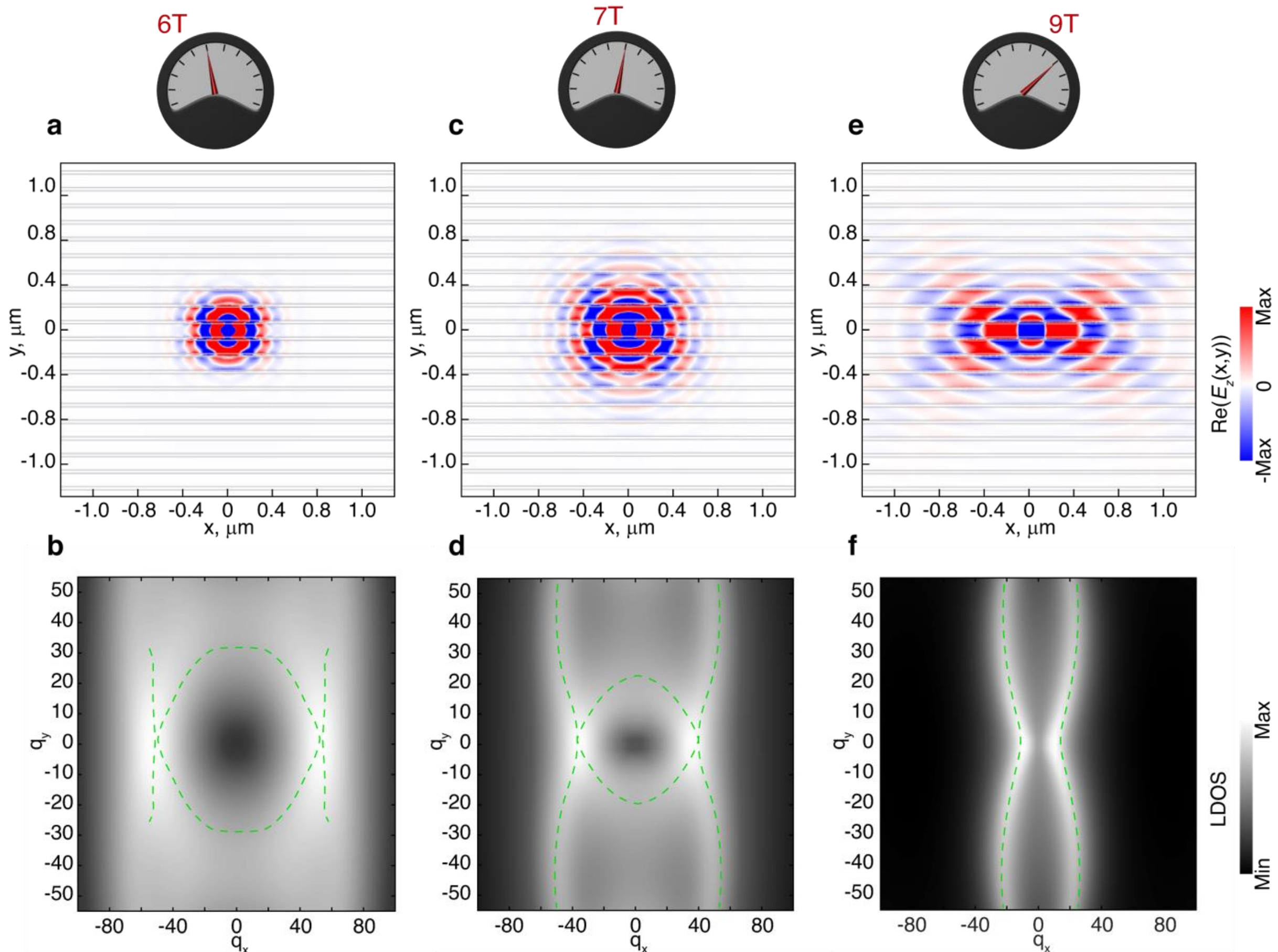

**Figure 2.** Quantum hyperbolic magnetoexciton polaritons in the metasurface for different values of the magnetic field. **(a, c, e)** Simulated real part of the electric field $\mathrm{Re}(E_z)$, excited by a vertical point dipole on top of a metasurface, at $B = 6,7,9$ T, respectively. **(b, d ,f)** Color plots with their bright maxima representing the IFCs of the metasurfaces at the same magnetic fields as in **(a, c, e)**. The green dashed lines trace the local maxima extracted from the Fourier transforms of $\mathrm{Re}(E_z)$ shown in panels (a, c, e). In all panels, the frequency is $\omega = 25$ THz, the lattice constant is $L = 150$ nm, and the ribbon width is $w = 120$ nm.

First, we study the behavior and topology of the IFC of the QHMEPs in our magnetized CNG metasurface. To this end, we perform full-wave numerical simulations where magnetized CNG nanoribbons, with a width of $w = 120$ nm, are arranged periodically with the lattice constant of $L = 150$ nm (see Methods). For each nanoribbon, we take the anisotropic conductivity obtained from the tight-binding calculations. A vertical electric point dipole placed 50 nm above the metasurface launches polaritons at frequency $\omega = 25$ THz. The real part of the spatial distribution of the vertical (out-of-plane) electric field, $\mathrm{Re}(E_z)$, is shown in **Figure 2a,c,e** for the applied magnetic field of $B = 6, 7$ and 9 T, respectively. It should be emphasized that, for our metasurface, the effective medium approach is,[26,27] strictly speaking, not valid as the criterium $q \ll G$ (where $G = k_0^{-1} 2\pi/L$ - reciprocal lattice vector) is not fulfilled. Therefore, the results presented here cannot be accurately reproduced using this approximation, which is commonly employed in metasurface studies (see Supplementary Note S5).

We observe a dramatic transformation of the polaritons' wavefront shape with increasing magnetic field, indicating a change in the polaritons' dispersion. At lower magnetic fields $B = 6$ T (**Figure 2a**), the wavefronts have an elliptical-like shape so that polaritons propagate along all the in-plane directions. In contrast, at higher magnetic field strength, $B = 7$ T (**Figure 2c**), the wavefronts stretch and start bending. The latter becomes more evident at the highest magnetic field strength shown, $B = 9$ T (**Figure 2e**), when the wavefronts acquire an inverted curvature and hyperbolic-like shape with the QHMEPs propagating predominantly along specific directions, forming features typical for the so-called "hyperbolic rays" determined by asymptotes of the hyperbolic IFCs.[25,26]

Fourier transforms of the simulated field distributions $E_z(q_x, q_y)$, qualitatively characterize the momenta distribution of QHMEPs. The maxima of $|E_z(q_x, q_y)|$ (shown with green dashed curves in **Figure 2b,d,f** for $B = 6, 7$ and 9 T, respectively) corresponding to the dominant field Fourier components of the excited QHMEPs, match well with the maxima of the local density of optical states (LDOSs) (also, see Supplementary Note S8 for direct comparison of full 2D Fourier transforms with LDOSs). The latter is represented by the grayscale colorplots in **Figure 2b, d, f**, and is calculated by summing up the amplitudes of the plane waves forming a Fourier-Floquet expansion of the electromagnetic field above the metasurface (see Methods).[28,29] The IFCs reveal a strong field confinement that corresponds to the large refractive index values, observed at the IFCs, and a clear evolution from closed (elliptical-like) to open (hyperbolic-like) topologies as the magnetic field increases (also, see Supplementary Note S4 illustrating the topological transition driven by tuning the frequency), confirming the high tunability of the polariton dispersion via external magnetic bias (also, see Supplementary Note S7 for details of the propagation properties of QHMEPs at different magnetic fields). In other words, variation of the applied magnetic field allows magnetically driven topological transition of the IFCs, which provides access to the previously unexplored hyperbolic quantum polaritons.

Furthermore, we notice that as the magnetic field increases, the IFCs become almost flat over a broad range of wavevectors, indicating the possibility of a so-called canalization regime - where polaritons propagate diffraction-less predominantly along a specific single spatial direction.[27,30,31] This regime allows channeling QHMEPs along a “virtual” naturally-emerging waveguide at the nanoscale, opening avenues for new possibilities of integrated flat quantum optics elements, particularly entanglement between quantum emitters via polaritons.[32–34]

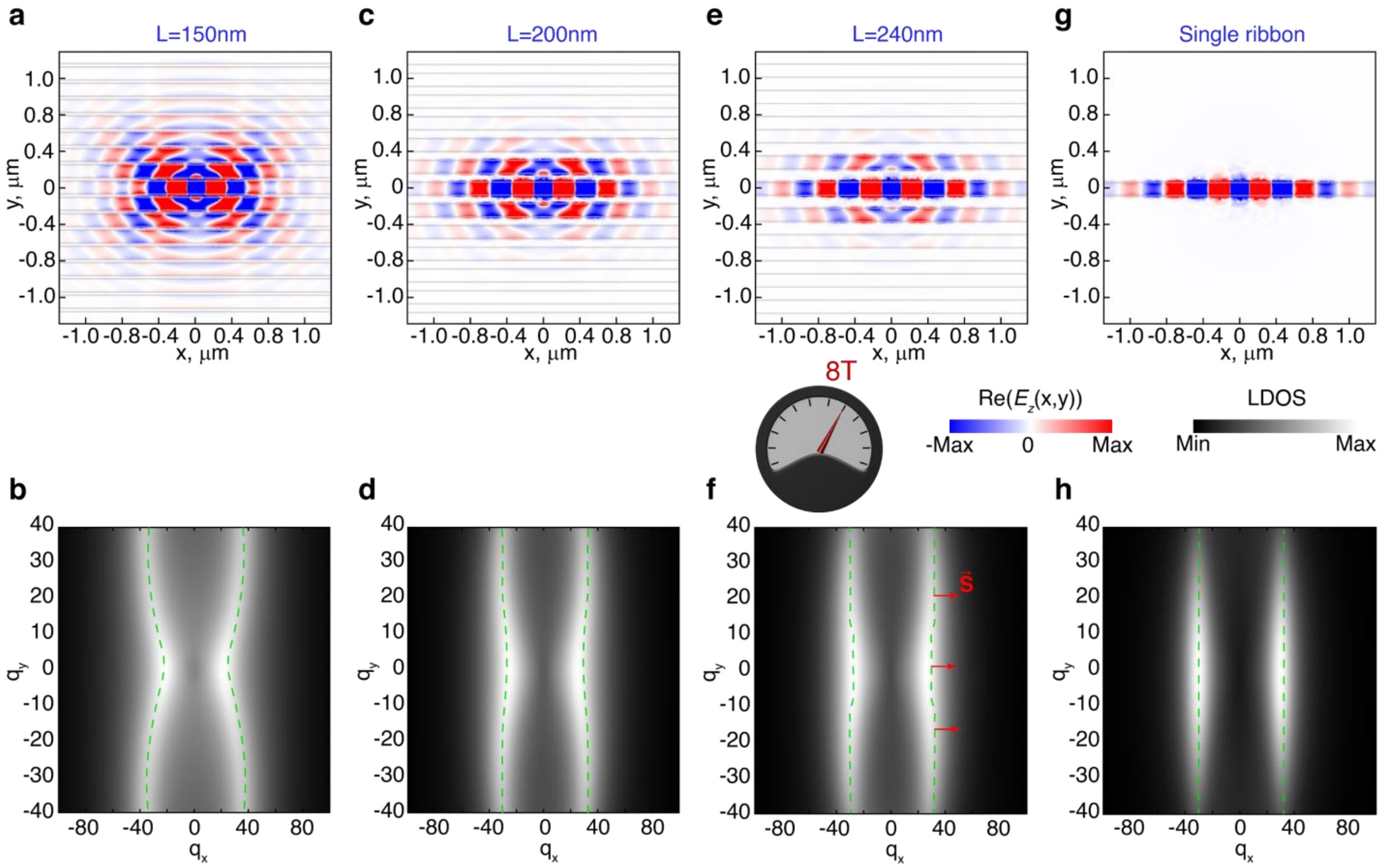

**Figure 3.** Tuning the topology of IFCs of QHMEPs by the metasurface geometry variation. **(a, c, e)** Simulated $\mathrm{Re}(E_z)$ of QHMEPs excited by a vertical point dipole on top of a metasurface with a lattice constant of $L = 150, 200, 240$ nm, respectively. **(b, d, f)** IFCs of the QHMEPs at metasurfaces with the same lattice constant as in **(a, c, e)**. **(g, h)** Simulated $\mathrm{Re}(E_z)$ exited in a single ribbon and its IFC, respectively. In all panels, the frequency is $\omega = 25$ THz, magnetic field is $B = 8$ T, and ribbon width is $w = 120$ nm.

To investigate the potential for the realization of the canalization regime in our metasurface, we focus on the parametric range in which the IFC flattens significantly. We first set the applied magnetic field to $B = 8$ T and the excitation frequency to $\omega = 25$ THz, while varying the period of the nanoribbon array, $L$, to explore its influence on the propagation characteristics of QHMEPs. **Figure 3a, c, e** show the simulated spatial distributions of $\mathrm{Re}(E_z)$ excited by a vertical point dipole in the metasurface with lattice constants $L = 150, 200, 240$ nm, respectively. For the smallest period $L = 150$ nm (**Figure 3a**), the field pattern of QHMEPs reveals concave-shaped wavefronts, which are corroborated by the corresponding hyperbolic Fourier transforms maxima and LDOS (green line and colorplot in **Figure 3b**, respectively). As the period increases to 200 and 240 nm (**Figure 3c** and **3e**, respectively), the field profile experiences a notable transformation: the QHMEPs become collimated along a single direction – along the ribbons – indicating the onset of canalized propagation. This behavior is also evidenced in the corresponding IFCs (**Figure 3d** and **3f**), which evolve into two nearly parallel lines. The flatness of the IFCs implies that the Poynting vector of QHMEPs (shown by the red arrows in **Figure 3f**) is perpendicular to the IFC at every point, facilitating diffractionless energy propagation along the ribbons.

Interestingly, our results are contrasting to the previously reported canalization for hyperbolic phonon polaritons (in a similar ribbon metasurface), which is directed across the ribbons.[27] On the one hand, this could be understood by the sign of the 2D diagonal conductivity tensor elements within the effective medium approach:[26,27] while in[27] $\mathrm{Im}\left(\sigma_{\parallel}^{eff}\right) < 0$ (along the ribbons) and $\mathrm{Im}\left(\sigma_{\perp}^{eff}\right) > 0$ (across the ribbons), in our metasurface $\mathrm{Im}\left(\sigma_{\parallel}^{eff}\right) > 0$ and $\mathrm{Im}\left(\sigma_{\perp}^{eff}\right) < 0$. On the other hand, the effective medium approach fails for our metasurface, as mentioned earlier. Thus, we assume that the origin of the QHMEPs canalization in our quantum metasurface emerges due to the coupling between high-$q$ polaritons in individual nanoribbons, which cannot be reproduced by the electrostatic approach.[26] Then, by changing the parameters of the metasurface, e.g., the distance between the nanoribbons, polaritons in individual nanoribbons decouple from each other. As a result, a point dipole source, depending upon its position, can launch QHMEPs along the corresponding single nanoribbon. To demonstrate our hypothesis based on the geometry of the metasurface, in **Figure 3g** we show the simulated $\mathrm{Re}(E_z)$ for a vertical dipole placed above a single magnetized CNG nanoribbon. The resulting excited QHMEP field profile resembles that observed in the metasurface with the largest period (**Figure 3e**). The corresponding IFC for the single nanoribbon (**Figure 3h**) is completely flat, featuring a single Poynting vector. The oscillation period $2\pi/k_0 q_x = 400$ nm observed in the metasurface along the ribbons matches with that of the waveguiding mode of a single ribbon and corresponds to the refractive index $\mathrm{Re}(q_x) \approx 30$ (**Figure 3f**), thus confirming our hypothesis regarding decoupling between QHMEP in the individual nanoribbons. The suggested interpretation of the physical origin of the canalization regime in our metasurface is also different from the one in naturally hyperbolic materials (such as $MoO_3$, $V_2O_5$, calcite, etc.).[35–37] Indeed, the canalization of polaritons in these van der Waals crystalline layers emerges without the need of nanostructuring, as it is dictated by the material properties of the slabs themselves.

In summary, we have demonstrated, for the first time, the emergence of quantum hyperbolic polaritons in metasurfaces composed of magnetized CNG ribbons. These highly confined QHMEPs exist at technologically relevant THz frequencies. By analyzing the IFCs and simulated near-fields of the QHMEPs, we revealed a magnetic-field-driven topological transition in the plasmonic dispersion – from elliptical to hyperbolic – and identified the onset of the canalization regime, in which energy propagates in a diffractionless, unidirectional manner. Furthermore, we showed that this canalization regime originates from effective decoupling between individual ribbons within the metasurface and thus can be realized through appropriate tuning of its geometric parameters. From a general perspective, our findings can be of a wide interest for the search of new types of collective excitations in emerging quantum materials.

## METHODS

### *Full-wave simulation*

We employ full-wave electromagnetic simulations using COMSOL Multiphysics (RF module). The dimensions of the computational domain are $5 \times 5 \times 1$ μm$^3$, with scattering boundary conditions applied to the outer domain boundaries. We model CNG ribbons as conducting sheets with anisotropic conductivity $\hat{\boldsymbol{\sigma}}$ obtained from the tight-binding calculation for different values of applied magnetic field and width of the ribbon. The structure is excited by a vertically oriented electric point dipole source, $\boldsymbol{d}$, placed 50 nm above the metasurface. The spatial distribution of the excited electric field (real part of the $z$-component of the field, $\mathrm{Re}(E_z)$) is visualized at the distance of 20 nm above the metasurface.

*Plane-waves expansion*

We represent the electromagnetic fields in the upper and lower half-spaces as infinite sums of plane waves. We match the fields at the metasurface interface through the boundary conditions, where the normalized conductivity tensor, $\hat{\boldsymbol{\alpha}}$, of the ribbons is expressed as a periodic function expanded in a Fourier series, $\hat{\boldsymbol{\alpha}}(\boldsymbol{r}) = \hat{\boldsymbol{\alpha}}(\boldsymbol{r} + n\boldsymbol{L}) = \sum_n \alpha_n e^{i\boldsymbol{G}_n \boldsymbol{r}}$, where $\boldsymbol{G}_n$ is a reciprocal lattice vector. This procedure yields an infinite system of coupled linear equations for the plane wave amplitudes.[28,29] To solve the system, we trunk it, ensuring convergence of the solution, for our metasurface, convergence is reliably achieved with a minimum of $n = 130$ retained modes. The obtained IFCs present the sum of the amplitudes of the Fourier harmonics of the reflected field $\sum_n |R_{pn}|^2$.

*Tight-binding calculation*

The electronic properties of a graphene ribbon are described by a tight-binding Hamiltonian that adopts only the nearest-neighbor hopping parameter $t = -2.8$ eV.[38] The effect of the orthogonal magnetic field, $\boldsymbol{B}$, is taken into account via the Peierls substitution $t_{mn} \mapsto t_{mn} \exp\left[\frac{2\pi}{\Phi_0}\int_{R_m}^{R_n} dr \cdot\right.$

$A(r)\Big]$, where $\Phi_0$ is the magnetic flux quantum, $\boldsymbol{R_n} = (x, y)$ denotes the atomic positions, and $A(r) = (-By, 0,0)$ is the vector potential in the Landau gauge. With translation symmetry being preserved along the x direction, the eigenvectors of the Hamiltonian can be labeled by the crystal wave number k along x, i.e., $\Psi_{nk}(y)$ for n being the band index. We characterize how much a normalized eigenvector distributes at the center of a 120 nm graphene ribbon by computing

$$r_{nk} = \sum_{-30\ \text{nm} \leq y \leq 30\ \text{nm}} |\Psi_{nk}(y)|^2$$

and show it as the color map in Fig.1b, where higher values of $r_{nk}$ are shown with yellow color and lower – with black. The optical conductivity is given by the Kubo-Greenwood formula

$$\sigma_{\mu\nu}(\omega, E_F, \beta) = \frac{2e^2\hbar}{iA_{\text{rib}}} \sum_k \sum_{m,n} \frac{\tilde{v}_\mu^{mn}(k)\tilde{v}_\nu^{nm}(k)}{E_m(k) - E_n(k) + \hbar\omega + i\eta} \times \frac{f_\beta[E_m(k) - E_F] - f_\beta[E_n(k) - E_F]}{E_m(k) - E_n(k)},$$

where $A_{\text{rib}}$ is the area of the ribbon, $\beta = 1/k_B T$ is the thermodynamic beta, $k_B$ is the Boltzmann constant, $T$ is the temperature, $\eta$ is a phenomenological parameter characterizing the electron scattering processes, $E_F$ is the Fermi energy, and $f_\beta(\epsilon) = 1/\left(1 + e^{\beta\epsilon}\right)$ is the Fermi-Dirac distribution function. Additionally, $\tilde{v}_{\mu(\nu)}^{mn}(\boldsymbol{k})$ are matrix elements of the operator $\tilde{v}_{\mu(\nu)} = U^\dagger(k) v_{\mu(\nu)}(k) U(k)$ with $v_{\mu(\nu)}(k)$ being the velocity operator along the $\mu(\nu)$-direction and $U(k)$ being a matrix constituted by the eigenvectors of the tight-binding Hamiltonian. In this work, we choose $T = 1$ K and $\eta = 0.002$ eV, which corresponds to a relaxation time around 330 fs.[39]

ASSOCIATED CONTENT

Supporting Information. Additional information about optical transitions in a graphene armchair ribbon, optical conductivity of armchair and zigzag nanoribbons, topological transitions of IFC of QHMEPs, applicability of the electrostatic effective medium approach, influence of quantum effects on the polaritonic properties of nanoribbons of different width and propagation properties of QHMEPs. (PDF)

The following files are available free of charge.

Preprint: Kateryna Domina; Tetiana Slipchenko; D.-H.-Minh Nguyen; Alexey B. Kuzmenko; Luis Martin-Moreno; Dario Bercioux; Alexey Y. Nikitin. Tunable hyperbolic Landau-level polaritons in charge-neutral graphene nanoribbon metasurfaces. 2025, arXiv:2506.23786. arXive. https://doi.org/10.48550/arXiv.2506.23786 (30 Jun 2025)

AUTHOR INFORMATION

**Corresponding Author**

* **Alexey Y. Nikitin** - Donostia International Physics Center, 20018 Donostia-San Sebastian, Spain; IKERBASQUE, Basque Foundation for Science, 48009 Bilbao, Spain; https://orcid.org/0000-0002-2327-0164. Email: alexey@dipc.org.

**Authors**

**Kateryna Domina** - Donostia International Physics Center, 20018 Donostia-San Sebastian, Spain; Universidad del Pais Vasco/Euskal Herriko Unibertsitatea, 48940 Leioa, Spain; https://orcid.org/0000-0002-9000-021X

**Tetiana Slipchenko** - Instituto de Ciencia de Materiales de Madrid (ICMM), CSIC, 28049 Madrid, Spain; https://orcid.org/0000-0003-3918-0275

**D.-H.-Minh Nguyen** - Donostia International Physics Center, 20018 Donostia-San Sebastian, Spain; Advanced Polymers and Materials: Physics, Chemistry and Technology, Chemistry Faculty (UPV/EHU), 20018 San Sebastian, Spain; https://orcid.org/0000-0003-1498-367X

**Alexey B. Kuzmenko** - Department of Quantum Matter Physics, University of Geneva, 1211 Geneva, Switzerland; https://orcid.org/0000-0001-9574-6435

**Luis Martin-Moreno** - Instituto de Nanociencia y Materiales de Aragon (INMA), CSIC-Universidad de Zaragoza, 50009 Zaragoza, Spain; Departamento de Fisica de la Materia Condensada, Universidad de Zaragoza, 50009 Zaragoza, Spain; https://orcid.org/0000-0001-9273-8165

**Dario Bercioux** - Donostia International Physics Center, 20018 Donostia-San Sebastian, Spain; IKERBASQUE, Basque Foundation for Science, 48009 Bilbao, Spain; https://orcid.org/0000-0003-4890-5776

**Author Contributions**

A.Y.N. conceived the study. The original ideas were discussed with L.M.-M and A.B.K. D.-H.-M.N. performed tight-binding calculation with supervision of D.B. T.S. contributed to the initial design of the structure and plane-waves expansion method for materials with anisotropic conductivity. K.D. performed the full-wave simulations and plane-waves expansion calculation. All authors took part in the discussion and interpretation of the results. A.Y.N. supervised the project. K.D. and A.Y.N. wrote the manuscript with inputs from all authors.

**Funding Sources**

The study was funded by the Department of Science, Universities and Innovation of the Basque Government (grant no. PIBA-2023-1-0007) and the IKUR Strategy; by the Spanish Ministry of Science and Innovation (grants no. PID2023-147676NB-I00, PRE2021- 097126, PRE2020-092758, CEX2023-001286-S). The research of A.B.K. was funded by the Swiss National Science Foundation (grant no. 200021-236697). D.H.M.N, and D.B. knowledge the support from the Transnational Common Laboratory Quantum – ChemPhys. LMM acknowledges the Government of Aragón through Project Q-MAD.

**Notes**

The authors declare no competing financial interest.

ABBREVIATIONS

2D, two-dimensional; CNG, charge-neutral graphene; QHMEP, quantum hyperbolic magnetoexciton polariton; IFC, isofrequency curve; LDOSs, local density of optical states.

For Table of Contents Use Only

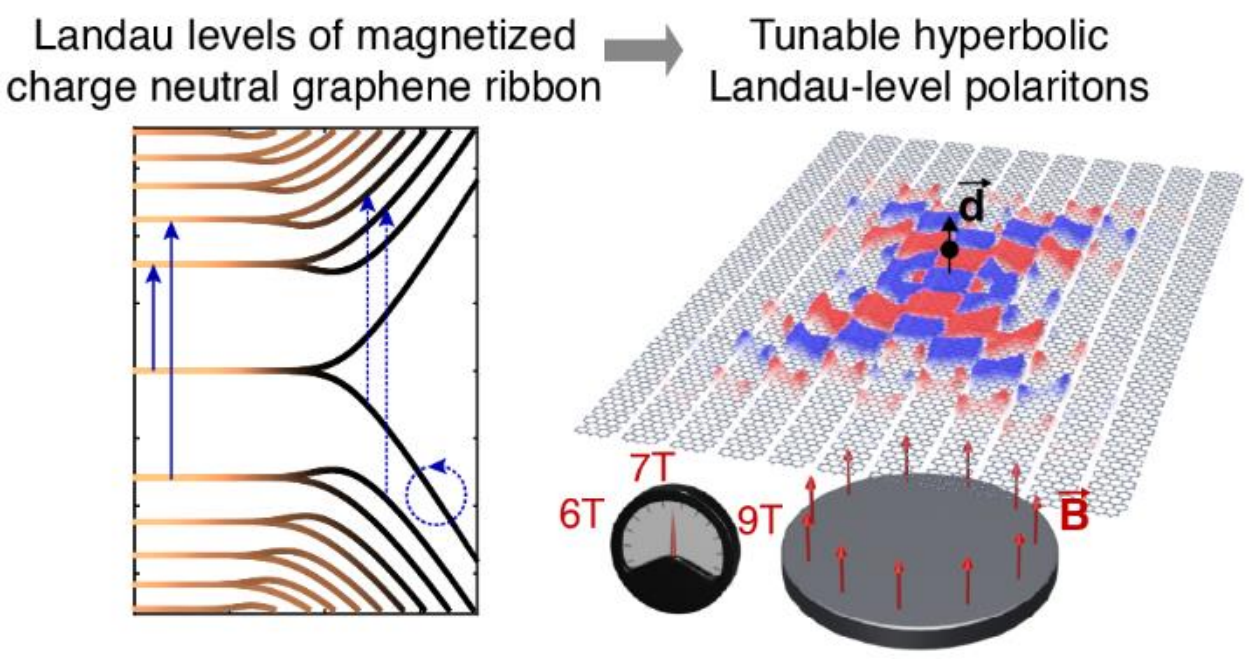